\documentclass[11pt,a4paper]{article}
\pdfoutput=1

\pdfoutput=1
\usepackage{jcappub}
\usepackage[utf8]{inputenc}
\usepackage{graphicx}
\usepackage{color}
\usepackage{xfrac}
\usepackage{verbatim}
 \usepackage{mathtools}
\newcommand{\bea}{\begin{equation}\begin{aligned}}
\newcommand{\eea}{\end{aligned}\end{equation}}
\def\lsim{\mathrel{\raise.3ex\hbox{$<$\kern-.75em\lower1ex\hbox{$\sim$}}}}
\def\gsim{\mathrel{\raise.3ex\hbox{$>$\kern-.75em\lower1ex\hbox{$\sim$}}}}

\providecommand{\f}[2]{\frac{{#1}}{{#2}}}

\newcommand{\ee}[1]{\begin{equation}#1\end{equation}}
\newcommand{\ea}[1]{\begin{align}#1\end{align}}

\title{Horizon Feedback Inflation}

\author[a]{Malcolm Fairbairn}
\author[b]{, Tommi Markkanen}
\author[c]{, David Rodriguez Roman}

\affiliation[a,c]{Department of Physics, King's College London\\Strand, London WC2R 2LS, United Kingdom}
\affiliation[b]{Department of Physics, Imperial College London,\\Blackett Laboratory, London, SW7 2AZ, United Kingdom}
                            
\emailAdd{malcolm.fairbairn@kcl.ac.uk}                            
\emailAdd{t.markkanen@imperial.ac.uk}
\emailAdd{david.rodriguez@kcl.ac.uk}

\abstract{We consider the effect of the Gibbons-Hawking radiation on the inflaton in the
situation where it is coupled to a large number of spectator fields. We argue that this will lead to two important effects - 
a thermal contribution to the potential and a gradual 
change in parameters in the Lagrangian which results from thermodynamic and energy conservation arguments.  
We present a scenario of hilltop inflation where the field starts trapped at the origin before slowly experiencing a 
phase transition during which the field extremely slowly moves towards its zero temperature expectation value.  
We show that it is possible to obtain enough e-folds of expansion as well as the correct spectrum of perturbations without hugely fine-tuned 
parameters in the potential (albeit with many spectator fields). We also comment on how initial conditions for inflation can arise naturally in this situation.}

\begin{document}
\begin{flushleft}
	\hfill		  KCL-PH-TH/2017-66\\
	\hfill		  IMPERIAL/TP/2017/TM/04 
\end{flushleft}
\maketitle

\section{Introduction}
Cosmological inflation is the leading paradigm which explains the horizon, flatness and defect problem of the extremely successful FLRW hot big bang model as well as explaining the source of the initial density perturbations observed to exist in the CMB \cite{Ade:2015lrj}.  Furthermore the exponential expansion seems to have an elegant explanation in field theory as being sourced by the potential energy of a field which rolls only slowly down to its minimum, its kinetic energy being red-shifted by this rapid expansion \cite{Linde:1981mu,Albrecht:1982wi,Starobinsky:1980te}.  It is also very difficult to think up alternatives to inflation which are natural, even with significant modifications of general relativity and those which do exist often create the wrong spectrum of perturbations \cite{Gasperini:1994xg}.

Unfortunately, there are several problems with the standard inflationary scenario.  The most difficult problem is (arguably) the fact that in order for inflation to start in the first place, one needs to find a Hubble patch in the early Universe across the entirety of which the potential energy dominates the kinetic energy of the field \cite{Vachaspati:1998dy}.  Another way of putting this is that in order to solve the horizon problem, one creates another one at earlier times.  Another problem is that in order for the kinetic energy to be redshifted by the expansion for long enough to explain the horizon and flatness problems, either the inflaton field has to be transplanckian during inflation or the shape of the potential below the Planck scale has to be extremely flat, in other words the extent in the scalar field direction has to be much larger in GeV than its height in the energy direction.

Cosmological inflation usually requires a period of de-Sitter or at least quasi de-Sitter expansion in order to obtain the many e-folds required to solve the horizon problem. The Gibbons-Hawking temperature associated with the cosmological horizon in de Sitter space \cite{Gibbons:1977mu}
\begin{equation}
T_H=\frac{H}{2\pi}
\,,\label{eq:TGH}
\end{equation}
plays a central role in inflation as it acts as the source of quantum fluctuations in the inflaton field which source density perturbations.  This thermal radiation is analogous to the thermal population observed by an accelerating observer i.e. Unruh radiation \cite{Unruh:1976db} and closely resembles the Hawking radiation which surrounds a black hole \cite{Hawking:1974sw}. In the case of a black hole, the thermal Hawking radiation escapes to infinity and energy is conserved only by postulating that the mass of the black hole is correspondingly released - a hypothesis which cannot be proved in semi-classical quantum gravity without including back-reaction but which does fit coherently into the theory of black hole thermodynamics \cite{Bekenstein:1974ax}. Black hole evaporation has an interpretation of resulting from a flux of particles with negative mass going into the black hole. A similar physical interpretation for de Sitter radiation is possible, where the Cosmological Constant is reduced from the addition of negative vacuum (zero-point) energy to the overall energy density of de Sitter \cite{Markkanen:2017abw}.

The author of \cite{Unruh:1976db} has famously said 'you could cook your steak by accelerating it' in regards to the physicalness of Unruh radiation \cite{Unruh:1989mw}. In light of the analogy between the Unruh and de Sitter radiation, assigning similarly physical characteristics to the radiation associated with the de Sitter horizon motivates its inclusion in the Einstein equations \cite{Padmanabhan:2003gd,Clifton:2014fja}. In a first principle approach this can be achieved in a prescription where the energy-momentum tensor sourcing gravity is coarse grained to include only the observable degrees of freedom, i.e. those inside the cosmological horizon \cite{Markkanen:2016jhg,Markkanen:2016vrp,Markkanen:2017abw}.

Including a source of continuous particle production in de Sitter space is well-known to bring about a qualitative change in the system's behaviour: when given enough time exponential expansion will cease implying that de Sitter space with particle production is not stable. In addition to thermal radiation from the horizon a de Sitter instability has been speculated to result from a variety other underlying mechanisms ranging from quantum gravity and infrared divergences in propagators to environment induced decoherence and the second law of thermodynamics \cite{Abramo:1997hu,Mukhanov:1996ak,Geshnizjani:2002wp,Tsamis:1992sx, Tsamis:1996qq,Tsamis:1996qm,Ford:1984hs,Polyakov:2007mm,Polyakov:2009nq,Akhmedov:2009be,Marolf:2010zp,Marolf:2010nz,Finelli:2011cw,ArkaniHamed:2007ky,Antoniadis:1985pj,Mottola:1984ar,Mottola:1985qt,Anderson:2013zia,Anderson:2013ila,Anderson:2017hts,Rajaraman:2016nvv,Markkanen:2016jhg,Markkanen:2016vrp,Markkanen:2017abw,Burgess:2015ajz,Dvali:2017eba,Anderson:2017hts}.

A de Sitter instability unavoidably leads to a dynamical cosmological constant and multiple studies have looked at the possibility that particle production could gradually reduce the value of the cosmological constant from a phenomenological point of view \cite{Freese:1986dd,Carvalho:1991ut,Lima:1994gi,Lima:1995ea,Alcaniz:2005dg,Wang:2015wga,Wands:2012vg}, see \cite{Gomez-Valent:2014rxa,Sola:2016jky} for observational investigations. It has been argued that since particle production from gravitational fields seems to be an irreversible process, this reduction is inevitable from a thermodynamical perspective \cite{Mottola:1985qt,Prigogine:1989zz,Calvao:1991wg,Lima:1995kd,Sekiwa:2006qj}.  Normally this effect is extremely small and would not really be a practical way of ensuring a small cosmological constant.  In particular the rate of decrease would not get rid of the cosmological constant fast enough in the late Universe to explain today's cosmic acceleration.  

It is interesting to consider the situation where the cosmological constant is not identified as a geometric term in the Einstein field equation, but is rather the energy density of a field which is located at a stable point in its potential where $dV/d\varphi=0$.  In this situation the same logic would imply that terms in the Lagrangian that set the scale of the potential would also decay over time due to the gravitational production of particles.  For example for the potential 
\begin{equation}
V(\varphi)=\frac{\lambda}{4}\left(\varphi^2-\varphi_0^2\right)^2\label{eq:pot}
\end{equation}
and for a field resting at the metastable origin $\varphi=0$, the particle production associated with the space-time curvature arising from the non-zero potential energy at the origin should have the effect of making $\varphi_0$ decrease with time. 

Another implication of the Gibbons-Hawking temperature would be the possibility that it might create non-negligible finite temperature contributions to the potential energy of scalar fields which are evolving in the quasi de Sitter background.

In thermal inflation, a thermal sector with the equation of state of radiation and coupled to the inflaton keeps the field trapped at the origin for a few e-folds until the thermal radiation has been redshifted away, decreasing its temperature \cite{Lyth:1995ka}.  If however the origin of the thermal radiation is the Gibbons-Hawking temperature associated with the horizon then its temperature will be constant and will not decrease over time if there is no corresponding change in the vacuum energy.  It is clear however that in this situation only a radiation bath with a large number of degrees of freedom, all of which are coupled to the scalar field, would be able to keep the field trapped at the origin.  Also once located and stablised there due to this horizon temperature, the field would essentially be stuck at the origin for all time, leading to a graceful exit problem. A constant $T_H$ would also violate the continuity equation \cite{Clifton:2014fja}.

If however $\varphi_0$ changes over time due to the production of radiation, one can imagine a situation where $\varphi_0$ and consequently the energy density at the origin, the rate of expansion of the Universe and the temperature of the thermal radiation all decrease with time.  The stable point at the origin then eventually becomes tachyonic, and field becomes free to roll away from $\varphi=0$, setting the initial conditions for inflation (see also \cite{Bastero-Gil:2016mrl} for a similar idea in a different context).  We argue that this could happen and that thermal effects would subsequently slow the phase transition sufficiently such that successful inflation can take place.  

In section \ref{sectwo} we will go through the equations of this scenario in more detail and study the dynamics of the field and how it might produce enough e-folds of expansion.  Then in section \ref{secthree} we will study the inflationary predictions, namely the perturbations and the spectral tilt, as well as comparing our analytic estimates to a numerical treatment.

\section{Phase transition with decaying vacuum energy \label{sectwo}}
Let us examine the situation with the potential (\ref{eq:pot}) in the presence of a thermal sector characterized by the Gibbons-Haking temperature (\ref{eq:TGH}) in more detail. We start by writing down the Friedmann equations
\ea{
\begin{cases}\phantom{-(}3H^2M_{P}^2&=T_{00}\equiv \rho
\\ -(3H^2+2\dot{H})M_{P}^2 &=T_{ii}/a^2\equiv p
\end{cases}\,.
\label{eq:e}}
Our model will consist of a scalar field $\varphi$ and importantly $N$ conformal fields that couple to $\varphi$ and are in thermal equilibrium with the horizon as described by (\ref{eq:TGH}). This will lead to and additional temperature component for the energy and pressure densities
\ee{\rho=\f{\dot{\varphi}^2}{2}+V(\varphi,T_H) + N\f{\pi^2}{30}T^4_H\,;\quad p=\f{\dot{\varphi}^2}{2}-V(\varphi,T_H) + \f{N}{3}\f{\pi^2}{30}T^4_H\,,}
and in addition to (\ref{eq:pot}) the potential contains a contribution from the thermalised conformal fields
\ee{
V(\varphi,T_H)=\frac{\lambda}{4}\left(\varphi^2-\varphi_0^2\right)^2+\frac{1}{2}Ng^2T_H^2\varphi^2\,.\label{eq:frida}
}
We will throughout work in the approximation where the energy density of the thermal component is subdominant to that of the potential and in particular to the vacuum  energy piece
\ee{\f{\lambda}{4}\varphi_0^4\equiv\rho_\Lambda\approx 3H^2M_{\rm P}^2\gg N\f{\pi^2}{30}T^4_H\,.\label{eq:conr}}
This condition will turn out to be easily satisfied for a large  parameter range i.e. $1 \gg N(H/M_{\rm P})^2(1440\pi^2)^{-1}$.

From (\ref{eq:e}) we get the dynamical Friedman equation of motion
\begin{equation}
-2\dot{H}M_{P}^2={\dot{\varphi}}^2+\frac{4}{3} N \frac{\pi^2}{30}T_H^4\,,
\label{FRW2}
\end{equation}
from which it is quite apparent that a thermal sector with the Gibbons-Hawking temperature is inconsistent with strict de Sitter space, but will lead to $\dot{H}<0$. Furthermore, it is easy to see that the continuity equation
\ee{\dot{\rho}+3H(\rho+p)=0\,,\label{eq:con}} can only be satisfied if $\varphi_0$ is not strictly constant but still provides the source for the continuous particle creation required for maintaining $T_H$ in the conformal fields despite the dilution from the expansion of space.
This one may understand from the situation when $\varphi=0$ giving $\rho\sim\rho_\Lambda$, but $\rho+p\sim T_H^4$. However as we will see we will not be relying on this feature of the current scenario in the current work other than to set initial conditions, the dynamics of $\varphi_0$ will be irrelevant by the time we come to calculate observables. Also we have deemed more natural to keep the usual interpretation of a strict coupling constant for the other constants of our theory, $\lambda$ and $g$, and only allow $\varphi_0$ change with time. Since the continuity equation (\ref{eq:con}) provides only one constraint in principle one could consistently allow also the other parameters to vary, at least from a purely phenomenological point of view.

From (\ref{FRW2}) we may conclude that the change in $H$ due to the thermal sector is very gradual: the first Hubble slow roll parameter for $\varphi=0$ reads
\begin{equation}
\epsilon=-\frac{\dot{H}}{H^2}=\frac{NH^2}{720\pi^2M_{P}^2}\label{eq:epsi} 
\end{equation}
and if initially $N\ll(720\pi^2 M_{P}/H_{init})^2$ then $\epsilon \ll 1$ is clearly satisfied.  For the parameters in the example situation that we will present later $\epsilon\simeq 10^{-7}$.
The evolution of the Hubble expansion rate as a function of the number of e-folds is then given by
\begin{equation}
\Big(\frac{H_{\text{\tiny SB}}}{H(N_e)}\Big)^{2}=1 + \frac{N_e \cdot N H_{\text{\tiny SB}}^2}{360\pi^2 M_{P}^2}
\end{equation}
where $H_{\text{\tiny SB}}$ is the value of the Hubble expansion at symmetry breaking.  (The number of e-folds $N_e$ are obtained simply by integrating $dN_e=Hdt$.) 

However, the Friedman equations also show for $\varphi=0$ that since $\rho=3H^2M_{P}^2\sim(\lambda/4)\varphi_0^4$, $\varphi_0$ will decrease slower than $T_H$. Therefore, the effective mass squared for $\varphi$
\begin{equation}
m^2\equiv Ng^2T_H^2-\lambda\varphi_0^2\,,\label{eq:meff}
\end{equation}
will eventually cross over to negative values, even if initially $Ng^2T_H^2\gg \lambda\varphi_0^2$. Simply put, at some point the system will undergo a phase transition. This phase transition in contrast to \cite{Lyth:1995ka} is extremely gradual. As we will show, it can take several thousands of e-folds to complete. This is due to the special nature of the thermal radiation as given by $T_H$: the thermal bath is continuously replenished by the decay of $\varphi_0$ and hence does not dilute in the usual fashion. 

After the phase transition $\varphi$ will start rolling to the new minimum and plays the role of the inflaton in the usual sense. The evolution of $\varphi$ can be characterized with (\ref{eq:meff}) to consist of three regions, $m^2\gtrsim 0$, $m^2\sim 0$ and $m^2\lesssim0$.
\subsection{$m^2\gtrsim 0$: gradual decay of $\rho_\Lambda$}
When the effective mass of the field (\ref{eq:meff}) is very large and positive the minimum of the potential is at $\varphi=0$ and supposing the field is in the minimum (\ref{eq:con}) leads (in the approximation that neglects $\dot{H}$) to
\ee{H=\f{ H_{init}}{\Big(\f{N H_{init}^3}{240\pi^2M_{P}^2}t+1 \Big)^{1/3}}~~\overset{t\rightarrow\infty}{\longrightarrow}~~\bigg(\f{240\pi^2}{N}\bigg)^{1/3}(M_{P}^2/t)^{1/3}\,,\label{eq:Hev2}}
which is consistent with the requirement
\ee{\dot{\rho}_\Lambda=-3H(\rho+p)=-{N}\f{4\pi^3}{15}T_H^5\,,}
as given by the continuity equation (\ref{eq:con}). As discussed after (\ref{FRW2}) for most cases the change in the Hubble is very gradual which justifies neglecting the derivatives of $H$ from the right side of the Einstein equation. Also interestingly the late time limit of (\ref{eq:Hev2}) is independent of the initial Hubble rate indicating that the case $m^2\gtrsim 0$ exhibits late time attractor evolution which is independent of initial conditions: after a sufficiently long time with $m^2\gtrsim 0$ the system will always relax to a configuration with $H\sim (M_{P}^2/t)^{1/3}$ and $\varphi=\dot{\varphi}=0$. For the remaining analysis we will choose the attractor configuration as our initial condition.

\begin{figure}
\begin{center}
\includegraphics[width=0.65\textwidth]{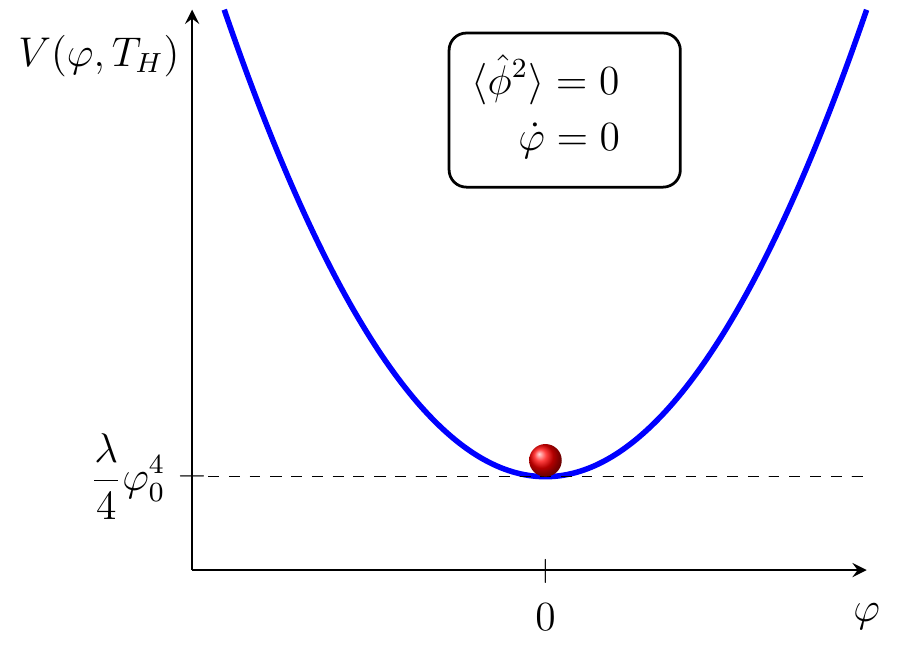}
\end{center}
\caption{\label{fig:BSSB} \it In the unbroken phase the field is at rest at $\varphi=0$ with practically no quantum fluctuations, $\langle\hat{\phi^2}\rangle\sim0$. The suppression of the quantum fluctuations is caused by the large effective mass making the field heavy with respect to the background curvature. }
\end{figure}

What is apparent from this section is that when $m\gtrsim0$ the system has fairly unremarkable behaviour: the field sits put in its vacuum state and the vacuum energy $\rho_\Lambda$ gradually decays. The quantum fluctuations around the mean we will denote as $\hat{\phi}$. When the field is heavy $m\gtrsim0$ they are similarly suppressed
\ee{\langle\hat{\phi^2}\rangle\sim0\,.} The case $m\gtrsim0$ is illustrated in Fig. \ref{fig:BSSB}.

As discussed after (\ref{eq:meff}) eventually the effective mass will vanish and the system will undergo a phase transition leading to interesting dynamics for $\varphi$.

\subsection{$m^2\sim 0$: large quantum fluctuations}
The phase transition happens when the effective mass of the field (\ref{eq:meff}) vanishes at the origin. The value of the Hubble parameter at and soon after the symmetry breaking transition is approximately given by the first Friedmann equation from (\ref{eq:e}) as
\begin{equation}
H_{\text{\tiny SB}}\simeq\sqrt{\frac{\lambda}{12}}\frac{(\varphi_{0})^2_{{\text{\tiny SB}}}}{M_{P}}\label{eq:HSSB}
\end{equation}
where $(\varphi_{0})^2_{{\text{\tiny SB}}}$ corresponds to the time of symmetry breaking. We can neglect both the thermal and kinetic contributions to the first Friedman equation and the kinetic term in the dynamical Friedman equation (\ref{FRW2}) so long as $\dot{\varphi}^2 \ll N T_H^4 \ll V$.  $N\gg \lambda/g^4$ may be obtained via (\ref{eq:conr}) and (\ref{eq:meff}) and it implies that $ N T_H^4 \ll V$ is fulfilled at the time of symmetry breaking. The second condition $\dot{\varphi}^2 \ll N T_H^4$ holds classically since at symmetry breaking $\dot{\varphi}=0$; but as it is showed later due to the quantum fluctuations $\varphi \neq 0$. As will be discussed more in section \ref{sec:CR} (see equation (\ref{eq:etinf0})) when the quantum fluctuations dominate over the classical motion one may write $|\dot{\varphi}|\sim H^2/(2\pi)$ which immediately implies $\dot{\varphi}^2 \ll N T_H^4$ for large $N$. 

\begin{figure}
\begin{center}
\includegraphics[width=0.65\textwidth]{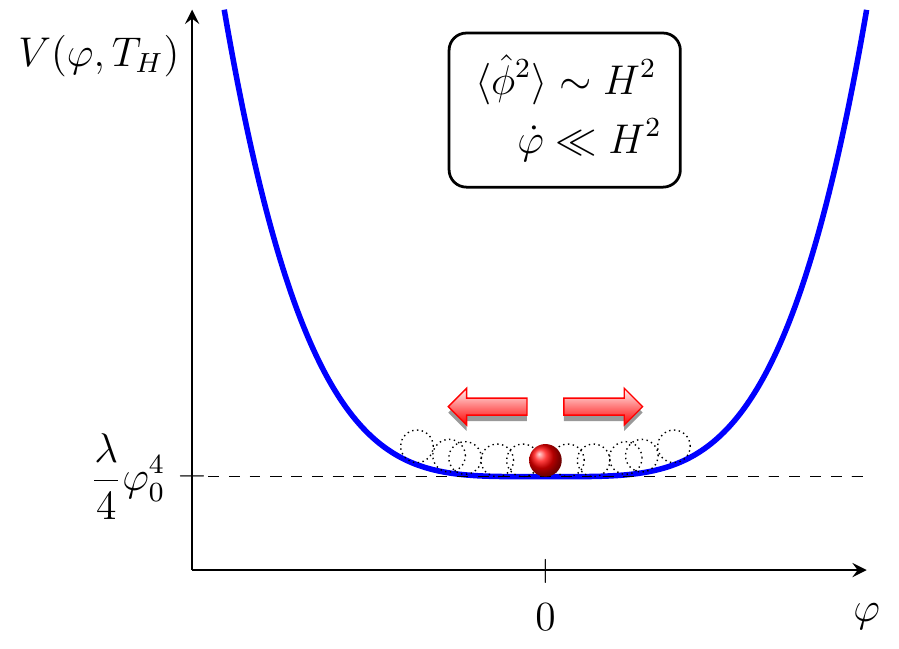}
\end{center}
\caption{\label{fig:ASSB} \it Close to symmetry breaking the classical (mean) field $\varphi$ is almost stationary, but the quantum fluctuations $\langle\hat{\phi}^2\rangle$ very large since the field is effectively light $m\sim0$.}
\end{figure}

However, importantly, $T_H$ and hence $m$ change very gradually and therefore immediately after symmetry breaking we expect a long epoch of very large quantum fluctuations before the rolling becomes dominant as illustrated in Fig. \ref{fig:ASSB}. The behaviour and magnitude of these fluctuations may be analytically solved via the stochastic formalism \cite{Starobinsky:1994bd,Starobinsky:1986fx}. This epoch of large quantum fluctuations is quite important in our model since as we will further discuss in section \ref{sec:CR}, when the $T_H$ has dropped enough the classical rolling will take over for which the initial condition will be dynamically set by the period of  quantum jumping.  We note that during this epoch where $m^2\sim0$ we would expect the large density perturbations to lead to the formation of domain walls, but since we expect to obtain many more than 60 e-folds of inflation after this period, we expect these evils to be swept outside the horizon.

In order to obtain a qualitative picture of the system's behavior as the representative value for the classical dynamics one may take the square root of the variance ${\varphi^2\equiv\langle\hat{\phi}^2\rangle}$.
Here we settle for solving this expectation value by using the Langevin equation in the Hartree approximation relegating a more complete discussion to section \ref{sec:num}. The relevant equation is \cite{Starobinsky:1994bd} 
\begin{equation}
\frac{d}{dt}\langle\hat{\phi}^2\rangle=\frac{H^3}{4\pi^2}-\frac{2m^2}{3H}\langle\hat{\phi}^2\rangle-\frac{2\lambda}{H}\langle\hat{\phi}^2\rangle^2.\label{eq:2pi}
\end{equation}
During this epoch there is little dynamics in ${\langle\hat{\phi}^2\rangle}$ and the two relevant limiting cases from the above are 
\ee{{\langle\hat{\phi}^2\rangle}\overset{m\rightarrow0}{=}\f{H^2}{\pi\sqrt{8\lambda}}\qquad{\rm and}\qquad{\langle\hat{\phi}^2\rangle}\overset{m\rightarrow-\infty}{=}\f{-m^2}{3\lambda}\,,\label{eq:sate}}
which are separated by the threshold
\ee{m^2_t\equiv\frac{3 H^2 \sqrt{\lambda }}{\sqrt{2} \pi }\,,\label{eq:thresh}}
i.e. for $-m^2\ll m_t^2$ ($-m^2\gg m_t^2$) the results coincide with those on the left (right) side of (\ref{eq:sate}). 

The solutions (\ref{eq:sate}) are approached only at the saturated limit when the system has been given enough time to equilibrate. The exact time this takes depends on the parameters of the potential. It is known that for a quartic theory the equilibration time scale in terms of $e$-folds is given by $1/\sqrt{\lambda}$  \cite{Hardwick:2017fjo,Grain:2017dqa}. If the time scale for the change in the Hubble rate in terms of $e$-folds or $\epsilon^{-1}$ is much longer than the equilibration time our approximation of using the results at the saturated limit is valid. With (\ref{eq:conr}), (\ref{eq:meff}) at $m\approx0$ and (\ref{eq:epsi}) we can write the condition $\epsilon^{-1}\gg \lambda^{-1/2}$ as
\ee{\f{15g^4N}{4\pi^2\sqrt{\lambda}}\gg 1\,,\label{eq:epsc}}
which again is easy to satisfy for large $N$.
Also, in our model the scale of symmetry breaking (\ref{eq:HSSB}) can be tuned by choosing $\lambda$,$N$ and $g$ in our potential (\ref{eq:frida}) with smaller scales corresponding to a slower dynamics as is evident from (\ref{eq:epsi}). Hence the condition in (\ref{eq:epsc}) can also be understood to imply the freedom to choose $H_{\text{\tiny SB}}$ low enough such that the saturated expressions in (\ref{eq:sate}) are a good approximation. 

In the absence of the horizon entropy thermalising a large number of fields coupled to the scalar field, this kind of symmetry breaking would not lead to good inflationary initial conditions since the Kibble mechanism would lead to large fluctuations from horizon to horizon with large field gradients that would prevent inflation from starting in the first place.  In this scenario, the thermal corrections to the potential prevent the kinetic energy dominated regions from running straight down to $|\varphi|=\varphi_0$.  Eventually therefore as the thermal dissipation continues to gently facilitate the slow phase transition and the field's classical position and motion starts to dominate over its quantum fluctuations, a region should emerge somewhere where the field is coherent enough across several horizons such that the well known difficulties obtaining inflationary initial conditions are overcome.

\subsection{$m^2\lesssim 0$: classical rolling}
\label{sec:CR}
\begin{figure}
\begin{center}
\includegraphics[width=0.65\textwidth]{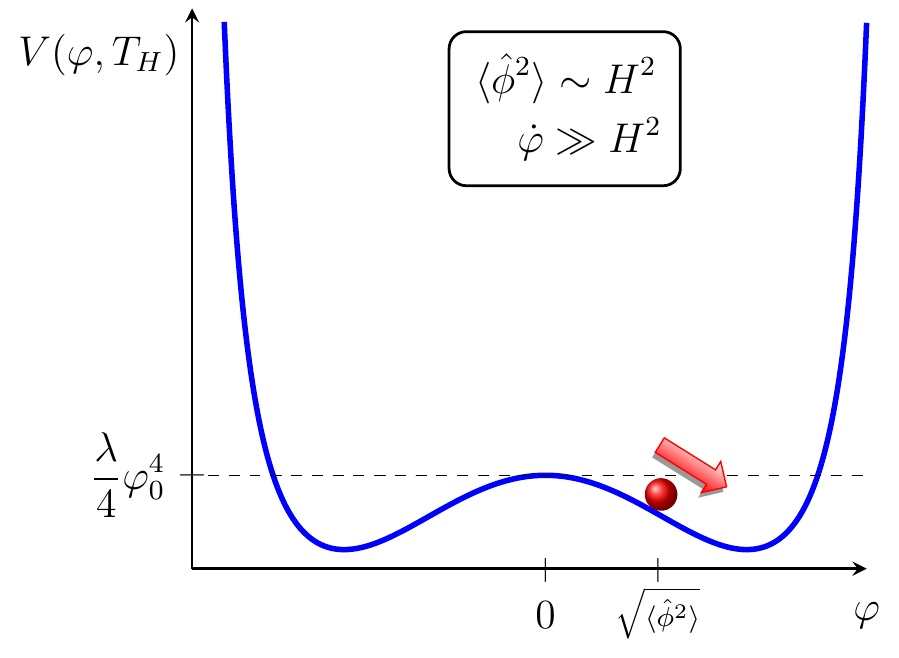}
\vspace{-1mm}
\end{center}
\caption{\label{AFSSB} \it The onset of classical rolling occurs when the steepness of the potential has increased enough. The initial value for the field is determined by the preceding random-walking epoch when $m\sim0$.}
\end{figure}
Strictly speaking one can only talk about a classical "rolling" once the minimum is further away from the origin than the stochastic vacuum expectation value that the field $\varphi^2\equiv\langle\hat{\phi}^2\rangle$ would obtain from the quantum fluctuations. As a first approximation one can say that the field starts rolling when in one e-fold the size of a single quantum jump ($H/2\pi$) is smaller that the distance traveled due to the classical rolling
\ee{|\dot{\varphi}|H^{-1}\gtrsim \f{H}{2\pi}\,,\label{eq:etinf0}}
which is the opposite to the usual condition for eternal inflation \cite{Guth:2007ng}.
More accurately we can use (\ref{eq:2pi}) for the dynamics of $\langle\hat{\phi}^2\rangle$.

As shown in the previous section, at the time of symmetry breaking ($m=0$), the variance of the field is approximately $2\frac{ \lambda}{H}\langle\hat{\phi}^2\rangle^2=\frac{H^3}{4\pi^2}$.  At times soon after symmetry breaking the mass term remains negligible $m^2\langle\hat{\phi}^2\rangle \ll 3 \lambda \langle\hat{\phi}^2\rangle^2$ and the field will remain constant at the value acquired at symmetry breaking until the mass is relevant, i.e. $-m^2- 3\lambda\langle\hat{\phi}^2\rangle=0$. At this point, the potential will be steep enough for classical slow roll to start and this classical motion will dominate over the quantum fluctuations (see Fig.\ref{AFSSB}).  Note that this condition agrees with the estimate (\ref{eq:etinf0}) since  $\frac{d}{dt}\langle\hat{\phi}^2\rangle=\frac{H^3}{4\pi^2}$. Once the classical rolling starts the mass is given by
\begin{equation}
-m^2\simeq 3 \lambda \langle\hat{\phi}^2\rangle=3\lambda \f{H^2}{\pi\sqrt{8\lambda}}\label{eq:etinf}
\end{equation}
and the value of the field and the speed are
\begin{equation}
\frac{d}{dN_e}\langle\hat{\phi}^2\rangle=\Big(\frac{H}{2\pi}\Big)^2 \qquad {\langle\hat{\phi}^2\rangle}{=}\f{H^2}{\pi\sqrt{8\lambda}}=\frac{-m^2}{3\lambda}
\end{equation}

Once classical rolling has been triggered, our model gives rise to the usual slowly rolling inflation. We emphasize that the initial conditions for inflation are set dynamically by the large quantum fluctuations prior to classical rolling and hence are not free parameters. Similarly, the start of slow roll is triggered dynamically once the potential has acquired sufficient steepness and occurs for a wide range of values for $\varphi_0$, in particular also for $\varphi_0/H\gg1$. Finally, we remind the reader that the neat attractor behaviour of the solutions prior to symmetry breaking (\ref{eq:Hev2}) also exhibit an independence from initial conditions. For these reasons we can conclude that the model presented here successfully evades all the usual fine-tuning issues of inflationary models.
\\
\section{Inflationary predictions\label{secthree}}
\label{sec:inf}
To have a successful inflationary scenario, we need enough e-folds (around 50-60) and we need to obtain the right perturbations  $\mathcal{P_R}$ and spectral tilt $n_s$ 50-60 e-folds before the end of inflation. As shown in the previous section, more than 60 e-folds of expansion can be obtained very easily in this scenario - the scale of inflation $H$ can be set to be much smaller than the Planck mass so $\epsilon\ll 1$ continues for many e-folds before $H$ differs significantly from the value at symmetry breaking $H_{\text{\tiny SB}}$, meaning that we get enough inflation.  Here we show how the perturbations are generated once the field is classically rolling to the minimum. 
The perturbations in the spatially flat gauge ($\Psi=0$) are defined as \cite{Baumann:2009ds}
\begin{equation}
\mathcal{R} = \frac{H \dot{\varphi}\,\delta \varphi}{\rho+p}\,,
\end{equation}
which we note will in our model lead to the usual expression encountered in models of warm inflation \cite{Berera:1995ie}.
The power spectrum takes the form
\begin{equation}\label{eq:pr2}
\mathcal{P}_{\mathcal{R}}=\Bigg(\frac{H\dot{\varphi}}{-2\dot{H}M_{P}^2}\Big(\frac{H}{2\pi}\Big)\Bigg)^2\,,
\end{equation}
\begin{figure}[!ht]
\begin{center}
\includegraphics[width=0.75\textwidth]{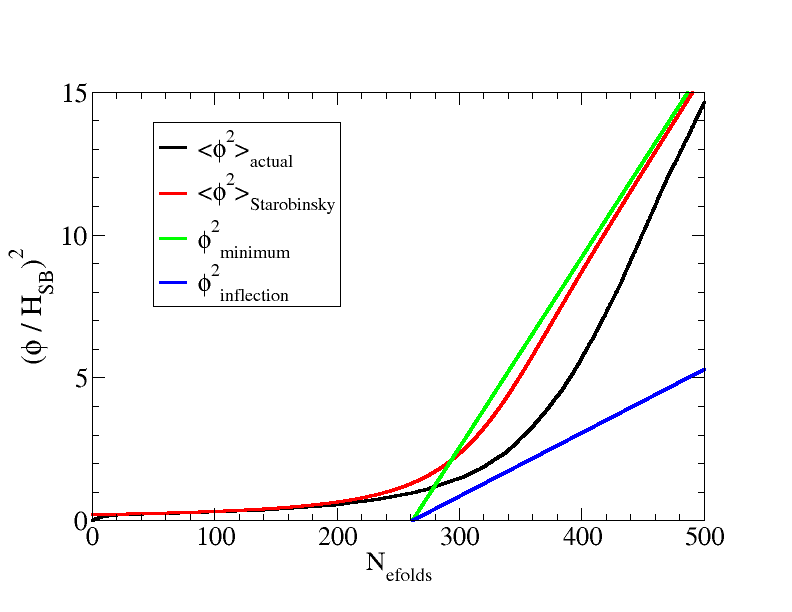}
\caption{\it Field dynamics with  $\lambda=10^{-2},N=10^{5.5},\varphi_0=10^{-1.41}M_{P} \text{ and } g=1$.  We plot the evolution of the field expectation vs. the Starobinsky estimate in units of $H_{\text{\tiny SB}}^2$ to show that they are different.  We also show the time evolution of the minimum and the inflection point (which the field must remain beyond in order to obtain a red spectrum). {$N_e=390$} is the epoch when the spectrum of perturbations and the tilt match the observed values by Planck.\label{fig:fields}}
\end{center}
\end{figure}
where from now on we set $\varphi^2\equiv\langle\hat{\phi^2}\rangle$. In contrast to single field slow roll inflation, in the current setting the perturbations and the tilt are given by
\begin{equation}
\mathcal{P}_{\mathcal{R}}=\Big(60\pi \frac{m^2 \varphi+\lambda \varphi^3}{N H^3}\Big)^2\,, \qquad n_s-1=6\epsilon -2\eta\label{eq:pert}\,,
\end{equation}
where we used the slow roll condition $\dot{\varphi}=-V'/3H$, and dropped the subdominant kinetic piece $\dot{\varphi}^2$ in (\ref{FRW2}). Furthermore, we note that $\epsilon$ is negligible in the calculation of the tilt. 

The difference with the usual single field slow roll inflation comes from the contribution of the thermal sector to the dynamical  Friedman equation (\ref{FRW2}), which both changes the vacuum energy decay and modifies the usual calculation for the spectrum because $\dot{H} \propto H^4$. 
Because of this, the perturbations follow the inverse of the usual $\mathcal{P}_{\mathcal{R}} \propto H^4/\dot{\varphi}^2$ behaviour.  This means that the usual expression for the tilt $n_s-1\propto \eta$ is not true and we get $n_s-1\propto-\eta$, where $\eta=M_{P}^2 \frac{V''}{V}$ is the usual slow roll parameter. So in order to obtain a red spectrum the inflaton needs to be evolving in the regime where the potential is convex ($V''>0$) i.e. beyond the inflection point at $V''=0$, contrary to the usual small field case.

There are a variety of different combinations of parameters that can give rise to the correct inflationary perturbations, however in what follows we will present a situation where the 50-60 e-folds of inflation we are interested in starts very soon after the symmetry breaking occurs.  In this situation, the value of $\eta$ is naturally small enough to give us the right tilt because soon after symmetry breaking $V''=0$ by definition. There is therefore no need to fine tune our tree-level parameters to give us a flat potential near the origin 
effectively bypassing the issue that the natural tree-level values are argued to give rise to $\eta \simeq 1$ \cite{Dine:1995kz}.

Our system has 4 free parameters $(N,\lambda, \varphi_0 \text{ and } g)$. However for the sake of clarity we choose here the final e-folds of inflation to occur soon after symmetry breaking and hence we can approximate the value of $\varphi_0$ during inflation by 
\begin{equation}
\frac{\varphi_0}{M_{P}}\approx\frac{(\varphi_0)_{\text{\tiny SB}}}{M_{P}}=\frac{4\pi\sqrt{3}}{\sqrt{N}g}\,,\label{eq:approi}
\end{equation}
which can be derived by making use of $m_{\text{\tiny SB}}=0$ in (\ref{eq:meff}) and (\ref{eq:HSSB}). With the above we effectively reduce the degrees of freedom from four to three. Also, if the observable part of the inflationary spectrum is going to take place soon after symmetry breaking then as soon as the classical rolling starts we want the field to be in the red tilt regime. For this to occur we need to ensure that $\frac{d}{dt}\langle\hat{\phi}^2\rangle>\frac{d }{dt} \frac{-m^2}{\lambda}$ at the inflection point, otherwise the minimum would move faster away from the origin than the field. If this were to occur, the spectral tilt would be blue (after symmetry breaking). By making use of very similar steps that led to (\ref{eq:approi}) and the Hartree approximation for $\langle\hat{\phi}^2\rangle$ from (\ref{eq:2pi}) one may show this condition to lead to the order of magnitude constraint $\frac{1}{g^2}<\frac{15}{4\pi^2}.$

So from now on, we will set for simplicity $g=1$. 
However we also emphasize that a red tilt can be obtained a long time after symmetry breaking, so $g=1$ is only a choice for the forthcoming calculation for the estimates of the perturbations and the tilt, and none of the above derivations relies on a specific value of g.

After this choice, we effectively have 2 degrees of freedom left ($N,\lambda$) that will determine the perturbations and the spectral tilt soon after the classical roll to the minimum starts.
It proves convenient to define a new parameter, $\alpha=\varphi/\sqrt{-m^2/3\lambda}$, where $\alpha=1$ means that the field is at the inflection point and $\alpha=\sqrt{3}$ means that the field is at the minimum of the potential. We can then approximate the perturbations and the tilt from (\ref{eq:pert}) and  with the help of equations from section \ref{sec:CR} by
\begin{equation}
\mathcal{P}_{\mathcal{R}}=\frac{2025}{\sqrt{2}\pi}\bigg(-\alpha+\frac{\alpha^3}{3}\bigg)^2 \frac{\sqrt{\lambda}}{N^2} \qquad n_s-1=\frac{-1+\alpha^2}{\sqrt{2}\pi}\sqrt{\lambda}
\end{equation}

It is interesting to note that the order of magnitude of the spectral tilt does not depend on $N$ while the magnitude of the perturbations depends on both ($\lambda \text{ and } N$). A smaller value of $\lambda$ will make our spectrum become more scale invariant and will allow us to reduce the number of conformal fields that need to be present to give us the right spectrum.  Hence we can shift the fine tuning between a small value of $\lambda$ and a large number of spectator fields.


Since in this model inflation takes place between the inflection point and the new minimum emerging due to broken symmetry, the field excursion of $\varphi$ scales as $\varphi\sim \sqrt{-m^2/\lambda}$. Furthermore, since for our choice $g=1$ this takes place soon after symmetry breaking where the mass parameter can be estimated as $-m^2\sim \sqrt{\lambda}H^2$, which with the Hubble rate from (\ref{eq:HSSB}) and (\ref{eq:approi}) indicates that for a large number of spectator fields the field excursions are sub-Planckian. For the parameters agreeing with observations this turns out to be true. This indicates that despite begin a small field model of inflation, the initial conditions are not fine-tuned but arise naturally as the favoured attractor solutions unlike what is usually encountered \cite{Goldwirth:1991rj}.

The approximations and estimates set out here agree with the numerical solution that we will turn to now.

\subsection{Numerical solution}
\label{sec:num}
In this section, we look at the numerical solution corresponding to the parameter choices we have made above. 

Before symmetry breaking the field lies at the origin $\varphi=0$ until $\varphi_0$ drops sufficiently for the potential at the origin to become tachyonic.  Classically of course, the field would then not move anywhere because $dV/d\varphi=0$ at the origin, and if we were to introduce a small perturbation away from $\varphi=0$ as an initial condition, then classically our final solution would depend strongly on this perturbation.  To get around this problem we evolve the field using the Langevin equation which takes into account the stochastic quantum fluctuations the field receives
\begin{equation}
\frac{d\varphi}{d N_e}=-\frac{m^2\varphi+\lambda \varphi^3}{3H^2}+ \frac{H}{2\pi}\xi
\end{equation}
where $\xi$ is a Gaussian white noise with zero mean and unit variance.  

For each combination of the remaining two free parameters $(N,\lambda)$ we perform $10^4$ numerical realisations of the Langevin equation and find the mean values of the magnitude of the perturbations $\mathcal{P_R}$ for each value of the tilt $n_s$ as the field evolves. These simulations are also used to verify that the dynamics of the field agrees with our analytical estimates from section \ref{sectwo}.

We know that to obtain a red spectrum ($n_s<1$) the field must lie between the inflection point and the minimum.

Figure \ref{fig:fields} shows how the variance of the field $\langle\hat{\phi}^2\rangle$ evolves with respect to the number of e-folds of expansion for the parameters $\lambda=10^{-2},N=10^{5.5}, \text{ and } g=1$ (making $\varphi_0=10^{-1.41}M_{P}$). Note that the initial epoch where we set the horizontal axis equal to zero is chosen arbitrarily.  We also plot in the same figure the variance of the field that one would expect using the Starobinsky-Yokoyama prescription for the variance in a perfect de Sitter spacetime corresponding to a vacuum energy equal to our evolving vacuum energy \cite{Starobinsky:1994bd}, assuming instant equilibration for the probability distribution.  The field expectation value lags behind this estimator, showing the importance of solving the Langevin equation.  The field approximately remains constant at this value until the minimum has dropped enough to make the potential steep and for it to possess an inflection point.  We also plot the position of the minimum as it moves out towards its zero temperature value and the position of the inflection point, showing that we remain on the good side to obtain a red spectrum.  Note that in a period of time during inflation corresponding to 60 e-folds of expansion, the variation of $\varphi_0$ is negligible relative to the variation in $\varphi$ so we assume that it is constant for calculational simplicity.

\begin{figure}
\begin{center}
\includegraphics[width=0.75\textwidth]{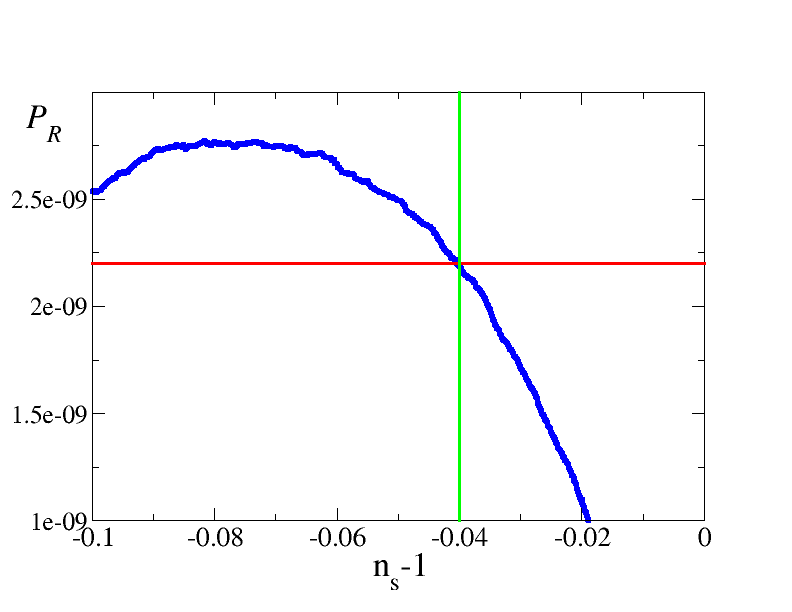}
\caption{ \it Amplitude of the spectrum of perturbations versus the tilt with  $\lambda=10^{-2},N=10^{5.5},\varphi_0=10^{-1.41}M_{P} \text{ and } g=1$. The blue line is the simulation (average over $10^4$ realisations) and the green and red lines are the cosmologically observed value of the spectrum and amplitude of pertubations respectively.\label{fig:spec}}
\end{center}
\end{figure}

Figure \ref{fig:spec} shows the average evolution of the magnitude of perturbations for the same parameters as a function of the spectral tilt.  We can see that these parameters can give rise to the correct combination of amplitude and spectrum for the perturbations to match what is observed in the Cosmic Microwave Background, i.e. $\mathcal{P_R}=2.2\times 10^{-9}$ and $n_s=0.96$ \cite{Hinshaw:2012aka,Ade:2015xua}.

The power spectrum of tensor fluctuations is given by the energy scale of inflation ( for example see equation 216 in \cite{Baumann:2009ds}), therefore for the parameters chosen, the tensor to scalar ratio that we get is $r=0.17$. This value is still within the 95 $\%$ CL from the Planck data  \cite{Ade:2015lrj} although lower values can be obtained by choosing a different set of parameters, since the scale of symmetry breaking can be easily tuned, eq. (\ref{eq:HSSB}).

In order to end inflation, we assume the temperature corresponding to the expansion rate during inflation falls below the mass of the particles which form the thermal bath affecting the potential for the scalar field.  We set this by hand to give us 60 e-folds after the epoch corresponding to the good values of $\mathcal{P_R}$ and $n_s$.

\section{Discussion/Conclusion}
In this work we have presented a possible new mechanism for inflation the early Universe.  We have argued that if one takes the Gibbons-Hawking temperature associated with the horizon of de Sitter space seriously, one is lead to a couple of conclusions that can affect the evolution of quantum fields in the early Universe significantly.  In particular we have argued that if there are enough fields coupled to the scalar field which takes the role of the inflaton, their Horizon induced temperature leads to thermal corrections to the potential which can affect its expectation value.  Since these thermal fluctuations do not redshift as rapidly as normal radiation, this effect can last for a significant number of e-folds of expansion.   

To summarise the mechanism - We consider a real scalar field with a $\mathbb{Z}_2$ symmetric potential with minima at $\pm\varphi_0$.  If the field starts at the origin, the non-zero energy density leads to de Sitter expansion and a cosmological horizon with an entropy.  The field is trapped at the origin due to its coupling to a large number of conformal spectator fields which have a non-zero temperature associated with this horizon entropy.  The height of the potential at the origin decays slowly as a result of the back-reaction of the thermal radiation on the parameter $\varphi_0$ in the Lagrangian.  The expansion rate therefore decreases as does the temperature $T_H$ of the thermal radiation until the mass at the origin becomes tachyonic, at which point a phase transition occurs and the field rolls away from the origin towards its zero temperature minimum at $|\varphi|=\varphi_0$.  This phase transition however occurs extremely slowly due to the same finite temperature corrections to the potential.  During this period of rolling we are able to obtain not only the correct number of e-folds but also the correct perturbations and spectral tilt as measured in the CMB.  This setup has two attractive features:-
\begin{itemize}
\item{Normally, for a field to act as an inflaton, it must have a super Planckian expectation value and/or very finely tuned parameters.  The mechanism we outline in this paper enables a potential which would otherwise not be flat enough to give rise to enough e-folds of inflation to do so due to the thermal effects of multiple spectator fields}
\item{The initial conditions for inflation may arise naturally as attractor solutions due to a slowly occurring phase transition. This happens despite the fact that inflation occurs with sub-Planckian values for the inflaton. This is the result of the parameters in the Lagrangian changing due to the back-reaction of the thermal radiation.}
\end{itemize}
It is clear that this scenario is not necessarily a panacea for all of the problems of inflation.  We need to assume that  parameters in the Lagrangian decay over time due to the back reaction of Hawking radiation at the horizon.  While there are many respected physicists who believe that such behaviour is probable and perhaps necessary, it is clear that further theoretical investigation and debate is required to put such speculation on a stronger footing.  We also require quite a large number of fields which are coupled to the inflaton in order to obtain the thermal braking required to obtain enough e-folds of inflation, in the example we have put forward here about $10^{5.5}$.  The masses of the fields need to be chosen such that inflation ends 50-60 e-folds after the epoch where the good perturbations and spectral index are set.  In return for this cost, we obtain a theory of inflation which doesn't require transplackian field excursions, does not require extremely small parameters in the Lagrangian (we use a value of $\lambda\sim 0.01$) and which may naturally explain the initial conditions for inflation.  Finally, it seems quite challenging to understand how the difference between this scenario and normal inflation could be distinguished experimentally.

\acknowledgments
This project was funded by the European Research Council through the project DARKHORIZONS under the European Union's Horizon 2020 program (ERC Grant Agreement no.648680). TM is supported by the STFC grant ST/P000762/1. The work of MF was also supported by the STFC.   


\bibliography{hawking}

\end{document}